\documentclass[times,twocolumn,final,nopreprintline]{elsarticle}
\usepackage{medima}
\usepackage{framed,multirow}
\usepackage{latexsym}
\usepackage{amsmath,amssymb,amsfonts}
\usepackage{graphicx}
\usepackage{url}
\usepackage{xcolor}

\usepackage{hyperref}

\definecolor{newcolor}{rgb}{.8,.349,.1}

\begin{document}

\verso{Abir Affane \textit{et~al.}}

\begin{frontmatter}

\title{The R-Vessel-X Project\tnoteref{tnote1}}%
\tnotetext[tnote1]{The research leading to these results was funded
by the French \emph{Agence Nationale de la Recherche} (Grant Agreement
ANR-18-CE45-0018).}

\author[IP]{Abir Affane}
\author[IP]{Mohamed A. Chetoui}
\author[LIRIS]{Jonas Lamy}
\author[CHU]{Guillaume Lienemann}
\author[CHU]{Rapha\"el Peron}
\author[IP]{Pierre Beaurepaire}
\author[LMR]{Guillaume Doll'e}
\author[IP]{Marie-Ange Lebre}
\author[CHU]{Beno\^it Magnin}
\author[CREATIS]{Odyss\'ee Merveille}
\author[IP]{Mathilde Morvan}
\author[LORIA]{Phuc Ngo}
\author[KITWARE]{Thibault Pelletier}
\author[LORIA]{Hugo Rositi}
\author[LMR]{St\'ephanie Salmon}
\author[KITWARE]{Julien Finet}
\author[LIRIS]{Bertrand Kerautret}
\author[CRESTIC]{Nicolas Passat}
\author[IP,cor]{Antoine Vacavant}

\cortext[cor]{Corresponding author: Antoine Vacavant;
  Tel.: +33-471099082; Email: antoine.vacavant@uca.fr}

\address[IP]{Universit\'e Clermont Auvergne, Clermont Auvergne INP, CNRS, Institut Pascal, F-63000 Clermont–Ferrand, France}
\address[LIRIS]{Universit\'e Lumière Lyon 2, CNRS, Ecole Centrale de Lyon, INSA Lyon, Universite Claude Bernard Lyon 1, LIRIS, UMR5205, 69007 Lyon, France}
\address[CHU]{Universit\'e Clermont Auvergne, Clermont Auvergne INP, CHU Clermont-Ferrand, CNRS, Institut Pascal, F-63000 Clermont–Ferrand, France}
\address[LMR]{Universit\'e de Reims Champagne Ardenne, CNRS, LMR, UMR 9008, Reims, France}
\address[CREATIS]{Univ Lyon, INSA‐Lyon, Universit\'e Claude Bernard Lyon 1, UJM-Saint Etienne, CNRS, Inserm, CREATIS UMR 5220, U1294,F‐69100, Lyon, France}
\address[LORIA]{Universit\'e de Lorraine, CNRS, LORIA, 54000, Nancy, France}
\address[KITWARE]{Kitware SAS, Villeurbanne, France}
\address[CRESTIC]{Universit\'e de Reims Champagne Ardenne, CRESTIC, Reims, France}

\received{xxx}
\finalform{xxx}
\accepted{xxx}
\availableonline{xxx}
\communicated{xxx}

\begin{abstract}
\textbf{1) Objectives:} This technical report presents a synthetic summary and the principal outcomes of the project R-Vessel-X (``Robust vascular network extraction and understanding within hepatic biomedical images'') funded by the French \emph{Agence Nationale de la Recherche}, and developed between 2019 and 2023. 
\textbf{2) Material and methods:} We used datasets and tools publicly available such as IRCAD, Bullitt or VascuSynth to obtain real or synthetic angiographic images. The main contributions lie in the field of 3D angiographic image analysis: filtering, segmentation, modeling and simulation, with a specific focus on the liver. 
\textbf{3) Results: }We paid a particular attention to open-source software diffusion of the developed methods, by means of 3D Slicer plugins for the liver anatomy segmentation (SlicerRVXLiverSegmentation) and vesselness filtering (SlicerRVXVesselnessFilters), and an online demo for the generation of synthetic and realistic vessels in 2D and 3D (OpenCCO). 
\textbf{4) Conclusion: }The R-Vessel-X project provided extensive research outcomes, covering various topics related to 3D angiographic image analysis, such as filtering, segmentation, modeling and simulation. We also developed open-source and free softwares so that the research communities in biomedical engineering can use these results in their future research. 
\end{abstract}


\end{frontmatter}

\section{Introduction}
\label{SEC:R-Vessel-X in Brief}

\subsection{Context}
\label{SSEC:Context}

Observing, analysing and/or modeling human vascular networks is a key-step for various medical applications, including e.g. diagnosis, surgery, drug delivery.
Medical image acquisition, such as X-ray Computed Tomography (CT) or Magnetic Resonance Imaging (MRI) plays a central role in this context.

Despite numerous developments in this field, computer extraction and modeling of vascular networks from 3-dimensional (3D) medical images is still a challenge, due to their high complexity and their partial visibility within most current medical acquisition systems, whatever the organ considered (brain, liver, heart, retina, \emph{etc.}).
Moreover, the reproducibility of such approaches is rarely addressed and experiments are often conducted on private datasets.

\subsection{Related Works}
\label{SSEC:Related Works}

After 30 years of research, vascular structure analysis from 3D images, that mainly relies on vessel segmentation, remains a complex issue, especially in the case of 3D images \citep{Lesage:MIA:2009,Moccia:CMPB:2018}.
Most efforts of 3D vessel segmentation were geared towards brain and heart vessels.
Nonetheless, the liver has also been under investigation, leading to various contributions. 
An exhaustive state of the art is beyond the scope of this article. 
The interested reader can refer to two recent surveys proposed by \cite{Ciecholewski:Sensors:2021} and \cite{Zhang:Entropy:2022} and the herein references.

In the context of liver vessel segmentation, most methods were designed to process X-ray Computed Tomography (CT) images, that constitute the state of the art for liver imaging.
During the period of the project (from 2019 to 2023), except few methods relying on classical image processing techniques (\emph{e.g.} graph cuts \citep{Guo:MBEC:2020}), a majority of the proposed approaches relied on deep learning paradigms: convolutional neural networks \citep{Kitrungrotsakul:CMIG:2019,Su:KBS:2021}, generative adversarial networks and autoencoders \citep{Cheema:TII:2021}, \cite{Yan:JBHI:2021} and multiscale paradigms \citep{Gao:TMI:2023}, graph neural networks \citep{Li:JBHI:2022}, deep supervision \citep{Hao:CMPB:2022} or transformers \citep{Wu:BMC-MI:2023,Wang:JBHI:2023}.

Although remaining less frequent, Magnetic Resonance Imaging (MRI) is becoming a relevant alternative to CT for liver image analysis \citep{Lebre:CBM:2019}.
A survey of artificial intelligence methods for MRI liver analysis was recently proposed by \cite{Hill:WJG:2021}.
During the last years, various strategies were considered.  \cite{Goceri:WJG:2021} relied on a classical clustering (K-means) approach.
\cite{Liu:CBM:2020} used convolutional neural networks to process MRI T2$^\star$ relaxometry.
\cite{Ivashchenko:MRI:2020,Guo:TMI:2022} took advantage of dynamic MR images to carry out classification-based analysis of structures including vessels.

\subsection{Objectives}
\label{SSEC:Objectives}

The purpose of the R-Vessel-X (``Robust vascular network extraction and understanding within hepatic biomedical images'') project was to develop novel solutions to extract vascular networks within 3D medical (CT or MR) images, with a specific focus on the liver.

\begin{figure}[!t]
\centering
\includegraphics[width=.9\linewidth]{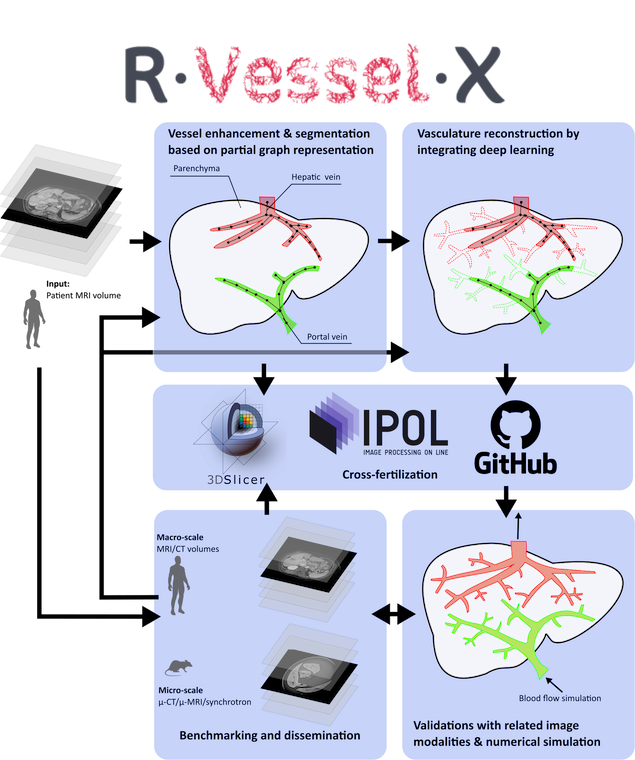}
\caption{Global flowchart and organization of the R-Vessel-X project.}
\label{fig:global}
\end{figure}

\begin{figure*}[!t]
\centering
\includegraphics[width=.9\linewidth]{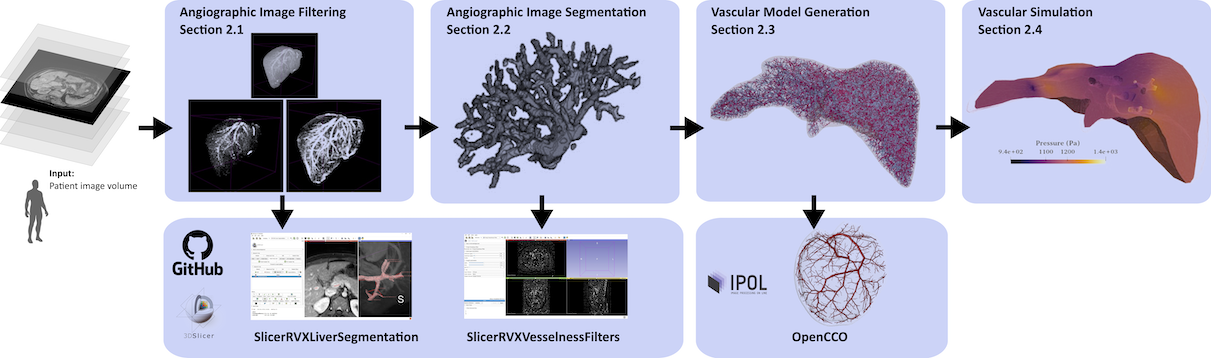}
\caption{Biomedical engineering workflow for angiographic image analysis, and associated organization of the paper.}
\label{fig:global2}
\end{figure*}

The scientific part of the project was organized as five interlinked work packages (WPs), visually summarized in Fig.~\ref{fig:global}.
The principal contributions of these WPs, together with pointers on related publications that provide more insight on the developed methods and tools, are given in the next section. 
The project outcomes were dedicated to various biomedical engineering topics related to 3D angiographic image analysis: filtering, segmentation, modeling and simulation, as depicted in Fig.~\ref{fig:global2}. 

\begin{figure*}[!ht]
  \begin{tabular}{ccc}
$\vcenter{\hbox{  \includegraphics[height=4cm]{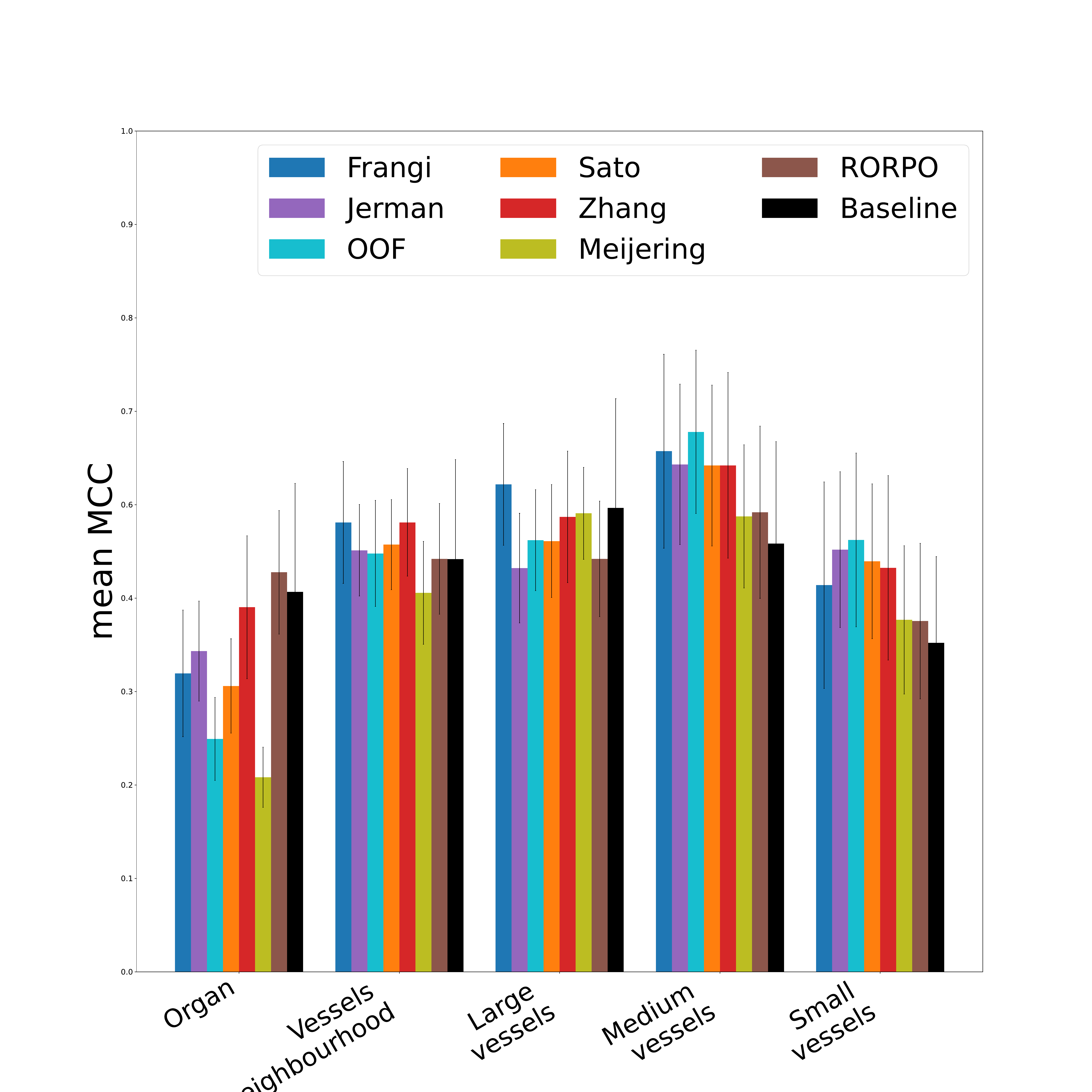}}}$&
$\vcenter{\hbox{  \includegraphics[height=4cm]{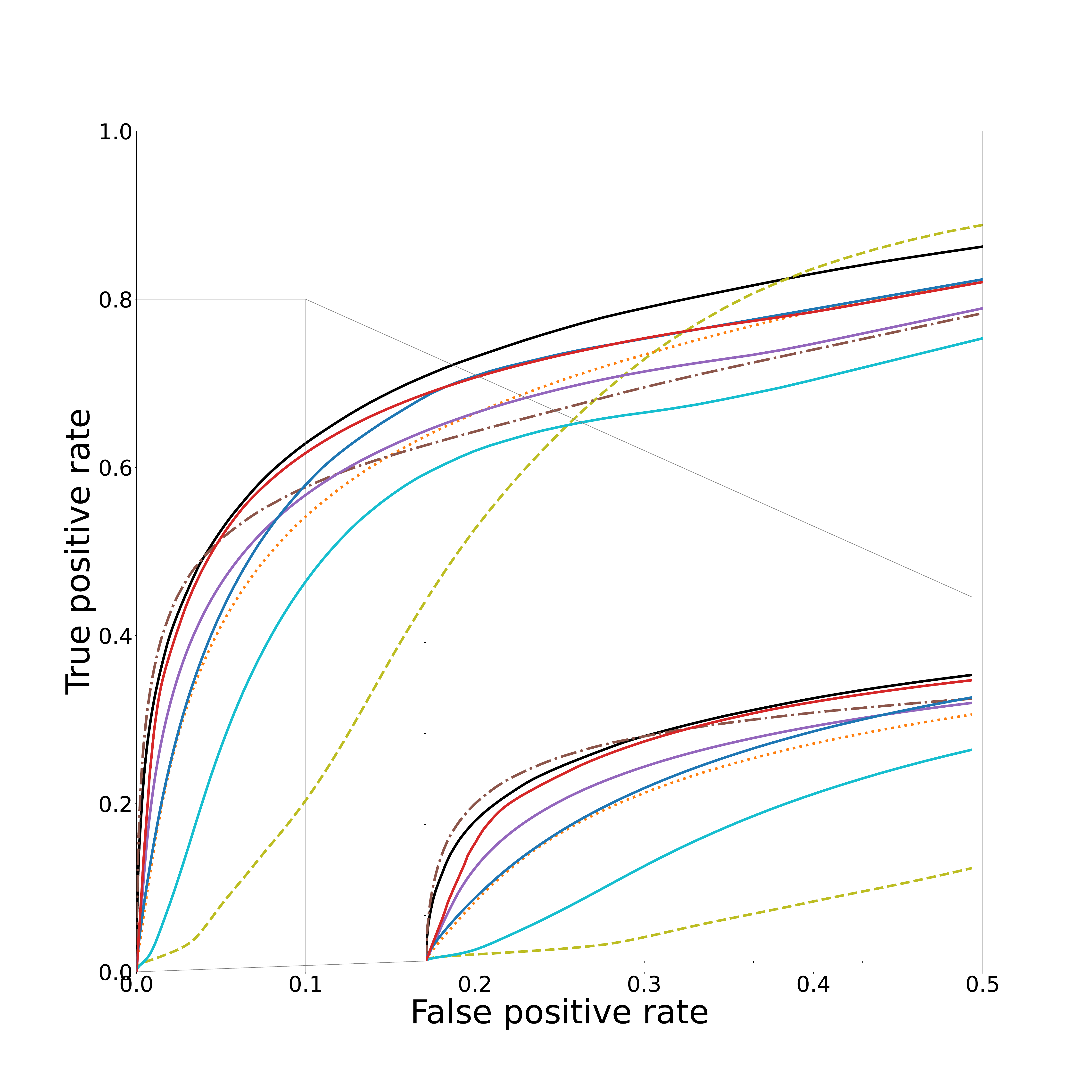}}}$&
$\vcenter{\hbox{  \includegraphics[width=0.5\textwidth]{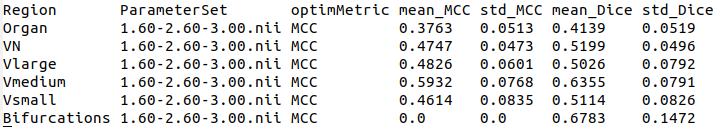}}}$\\
  (a) & (b) & (c) 
  \end{tabular}
  \caption{Typical output of the benchmark tools module illustrating numerical results obtained for a set of parameters and different measures.}
  \label{fig:bench_module2}
\end{figure*}

\begin{table*}[!ht]
\small
  \begin{center}
  \caption{Quantitative results (mean $\pm$ standard deviation) by regions / vessel size for the IRCAD dataset. More details can be found in~\citep{DBLP:journals/tmi/LamyMKP22, DBLP:phd/hal/Lamy23}.}
  \label{tab:VesselssizeIrcad}
  {    
    \begin{tabular}{lcccc}
           \hline 
           & Neighborhood &  Small & Medium & Large \\ 
           \hline 
 Baseline    &  $ 0.527 \pm 0.110 $          & $  0.597 \pm 0.136 $          & $  0.557 \pm 0.117 $ &  $  0.424 \pm 0.097 $ \\
 Frangi    &  $ \textbf{0.581} \pm 0.065 $ & $  \textbf{0.627} \pm 0.087 $ & $  0.660 \pm 0.113 $ &  $  0.506 \pm 0.118 $ \\
 Jerman    &  $ 0.521 \pm 0.060 $          & $  0.496 \pm 0.083 $          & $  0.622 \pm 0.110 $ &  $  0.525 \pm 0.104 $ \\
 Meijering &  $ 0.522 \pm 0.049 $          & $  0.669 \pm 0.044 $          & $  0.602 \pm 0.082 $ &  $  0.462 \pm 0.077 $ \\
 OOF     &  $ 0.556 \pm 0.051 $          & $  0.574 \pm 0.067 $          & $  0.681 \pm 0.097 $ &  $\textbf{0.559} \pm 0.096$  \\
 RORPO     &  $ 0.501 \pm 0.075 $          & $  0.504 \pm 0.080 $          & $  0.573 \pm 0.115 $ &  $  0.435 \pm 0.104 $ \\
 Sato    &  $ 0.542 \pm 0.057 $          & $  0.548 \pm 0.086 $  & $\textbf{0.629} \pm 0.105 $  &  $  0.522 \pm 0.091 $ \\
 Zhang     &  $ 0.551 \pm 0.074 $          & $  0.561 \pm 0.101 $          & $  0.619 \pm 0.126 $ &  $  0.497 \pm 0.124 $ \\
           \hline

            \end{tabular}  
  }
  \end{center}
\end{table*}

\section{Contributions}
\label{SEC:Contributions}

\subsection{Angiographic Image Filtering}
\label{SSEC:Angiographic Image Filtering}

\begin{figure}
\small
  \begin{tabular}{c@{\hspace{0.1cm}}c@{\hspace{0.1cm}}c}
    \includegraphics[clip = true, trim  = 0 50 0 80, width=0.15\textwidth]{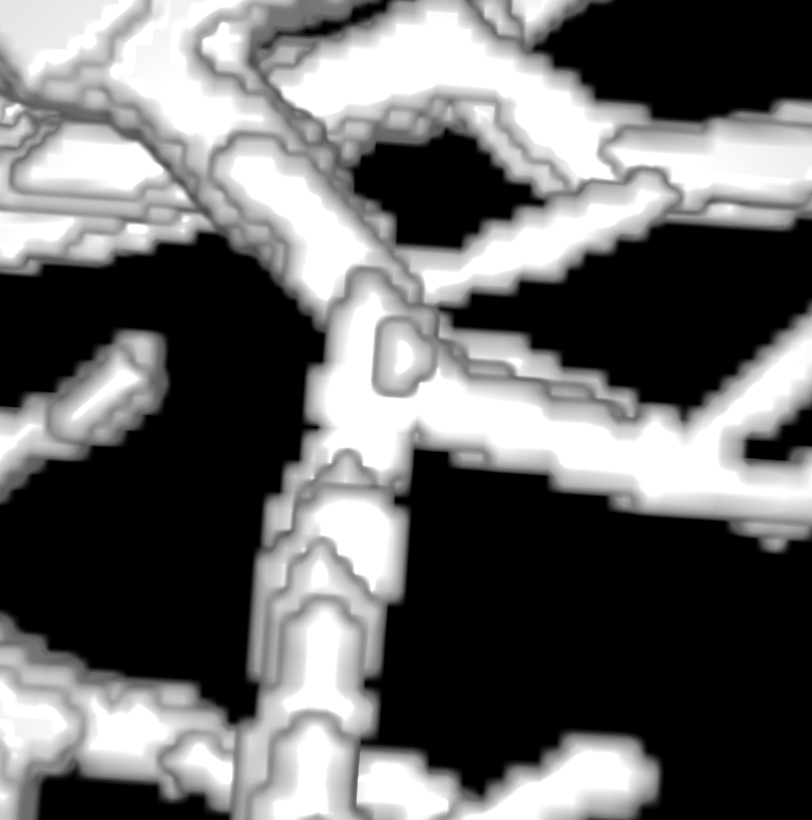}&
    \includegraphics[clip = true, trim  = 0 50 0 80, width=0.15\textwidth]{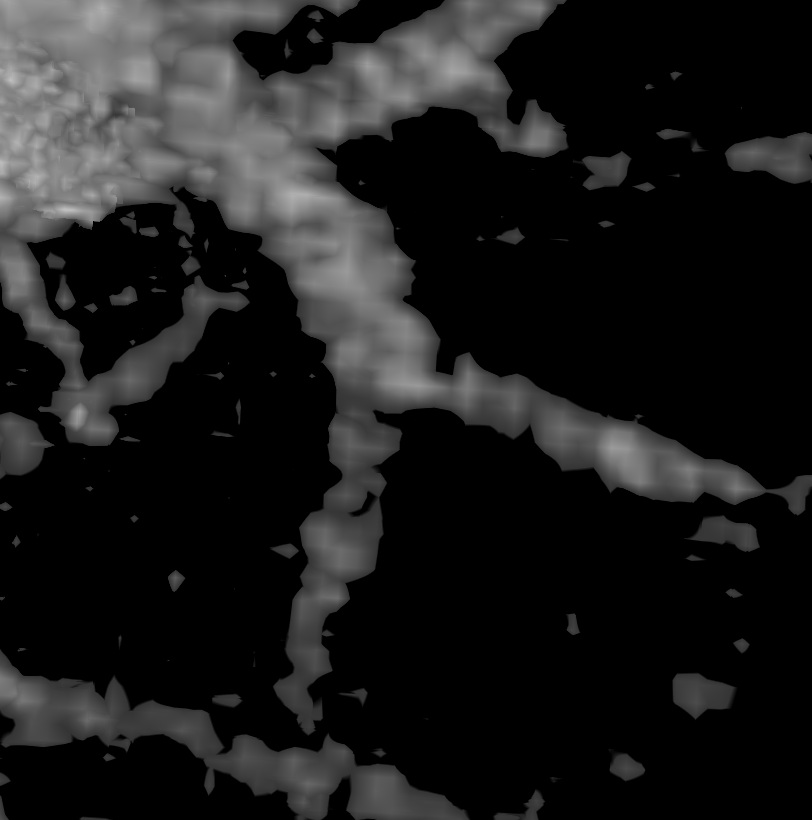}&
    \includegraphics[clip = true, trim  = 0 50 0 80, width=0.15\textwidth]{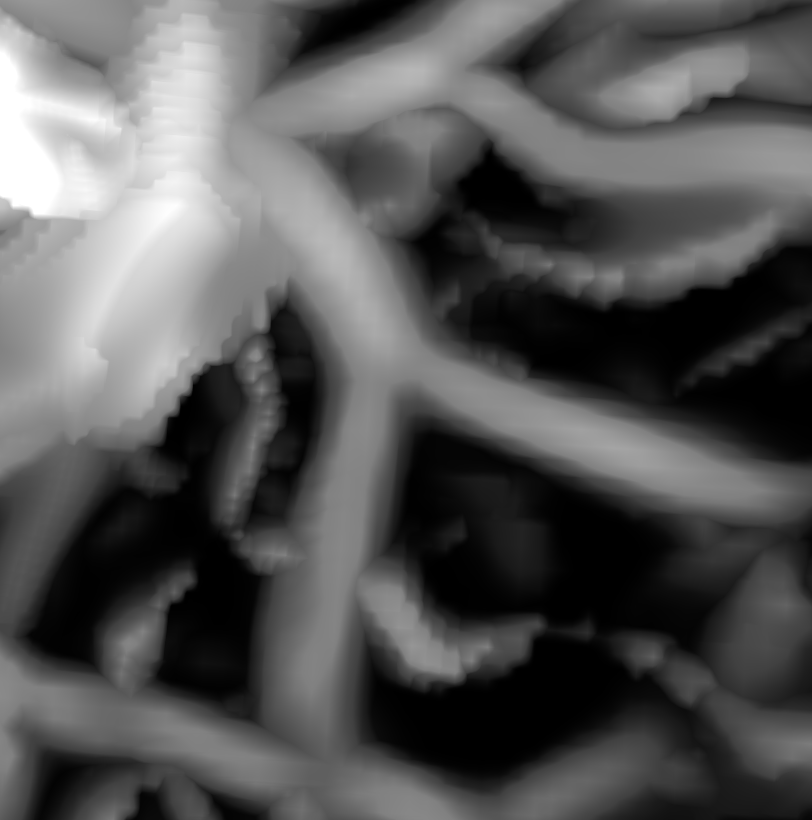 }\\
    (a) Ground-truth  & (b) Optimal threshold  & (c) Frangi \\
& from source intensity &    \small{\cite{DBLP:conf/miccai/FrangiNVV98}}\\
    \includegraphics[clip = true, trim  = 0 50 0 80, width=0.15\textwidth]{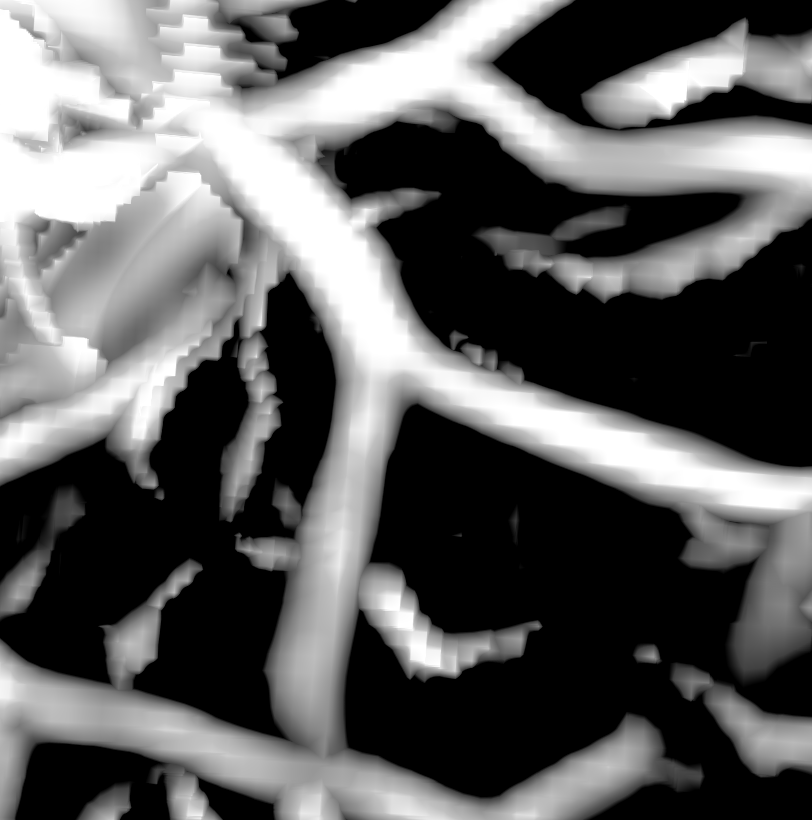}&
    \includegraphics[clip = true, trim  = 0 50 0 80, width=0.15\textwidth]{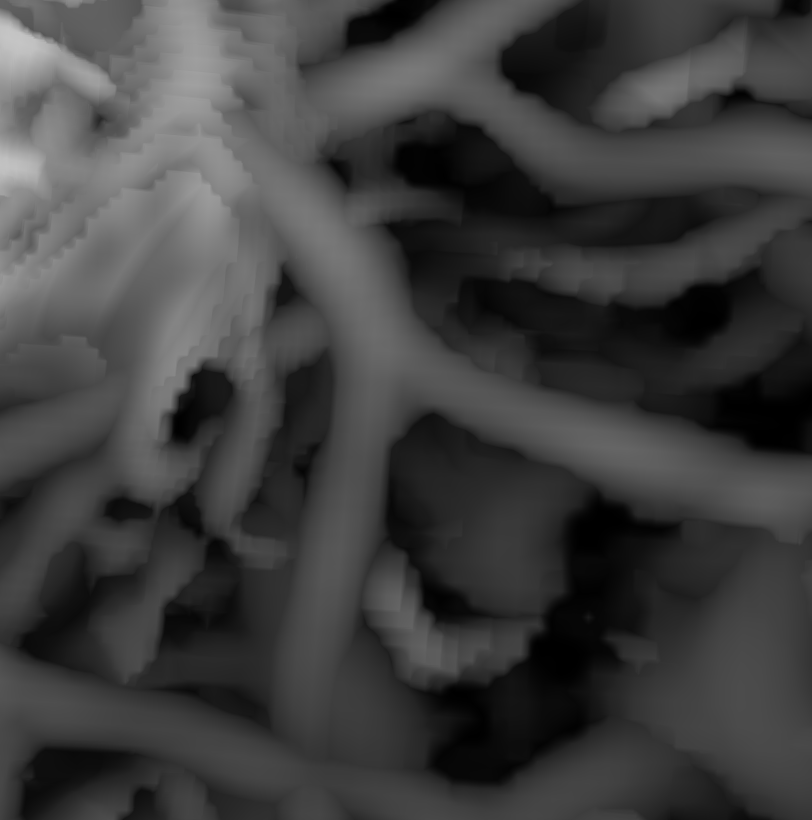}&
    \includegraphics[clip = true, trim  = 0 50 0 80, width=0.15\textwidth]{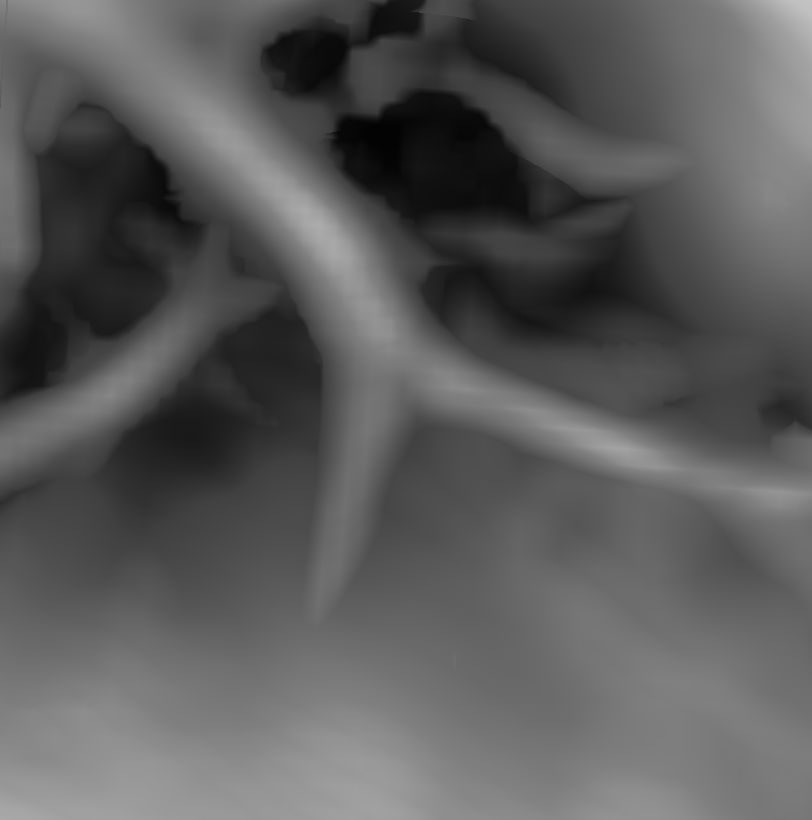}\\
    (d) Jerman & (e) OOF & Meijering \\
       {\small{\cite{Jerman:TMI:2016}}}   & {\small{\cite{Law2008_OOF}}}        &       {\small{ \cite{Meijering2004_neurite_vesselness}}}\\ 
    \includegraphics[clip = true, trim  = 0 50 0 80, width=0.15\textwidth]{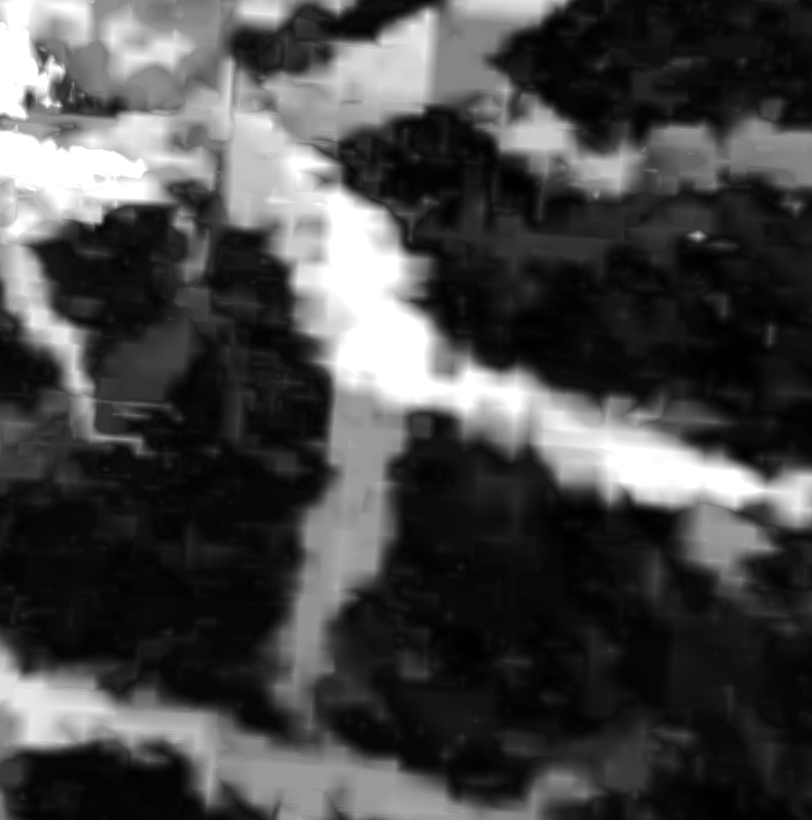 }&
    \includegraphics[clip = true, trim  = 0 50 0 80, width=0.15\textwidth]{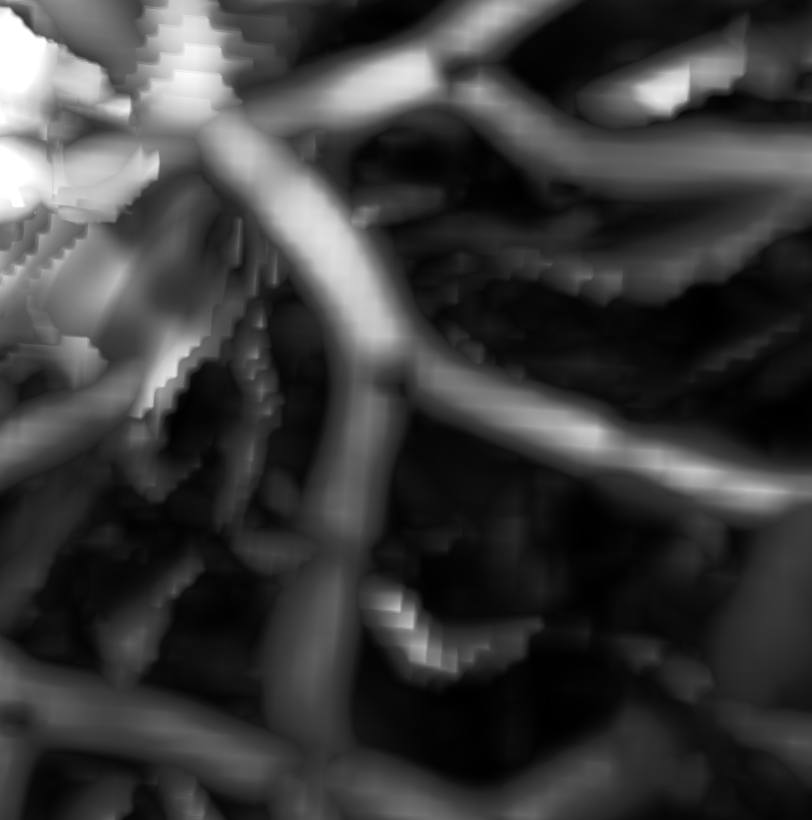 }&
    \includegraphics[clip = true, trim  = 0 50 0 80, width=0.15\textwidth]{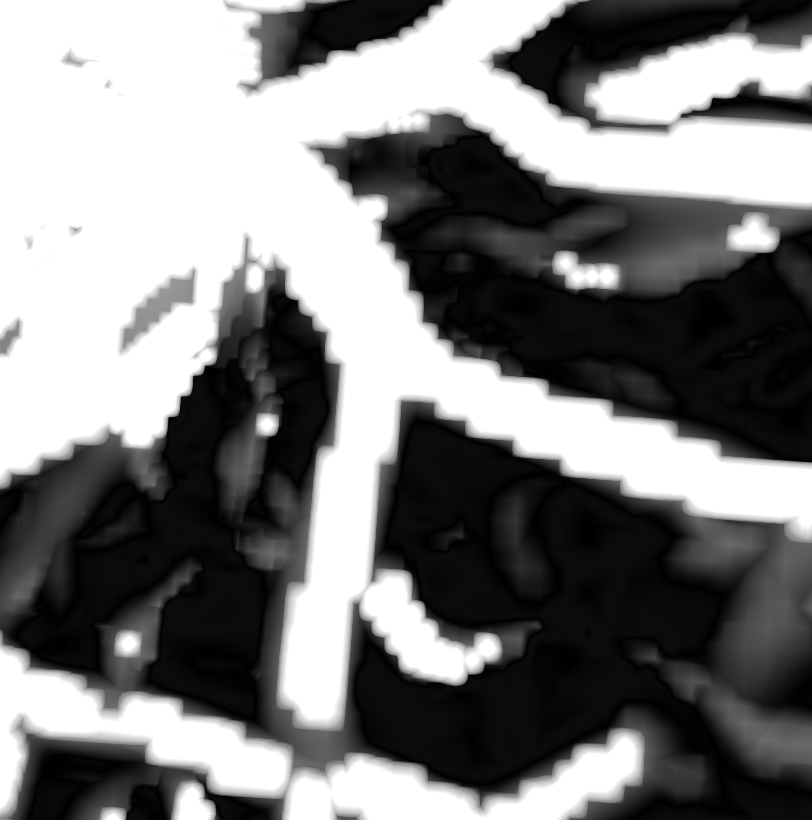}\\
    (g) RORPO & (h) Sato & (i) Zhang\\
         {\small{\cite{DBLP:journals/pami/MerveilleTNP18}}}     & {\small{\cite{Sato1998_vesselness}}}        &       {\small{ \cite{Zhang2018_liver_fuzzy_connectedness}}}\\ 
  \end{tabular}
  \caption{MIP visualisation of the filtering results  on a bifurcation area of the VascuSynth dataset.}
\label{fig:vesselCCOheart}

\end{figure}

Filtering with the purpose of denoising and/or enhancing the vascular signal vs. the remaining patterns has been one of the first research topics in angiographic image analysis.
Over the years, a wide range of filters were proposed, from derivative-based \citep{DBLP:conf/miccai/FrangiNVV98} to non-linear approaches \citep{DBLP:journals/pami/MerveilleTNP18}.
Despite a large choice of operators, filtering remains a complex task.
Indeed, the efficiency of a filter depends  on the type of acquisition and on the vascular structure being imaged.
Generally the proposed filters are also governed by many parameters that require ad hoc setting.
It follows that choosing the right filter with the right parameters is tricky, \emph{a fortiori} in absence of a general benchmarking framework.
Building such a framework has been the purpose of the PhD thesis carried out by Jonas Lamy \citep{DBLP:phd/hal/Lamy23}.
Its purposes were twofold: (1) designing a generic protocol that allows to run and fairly compare filtering operators with controlled parameters, and (2) providing reliable versions of the most popular angiographic filters.
This thesis finally led to revisiting seven gold standard filters, embedded in a C++ library.
They were compared on three public datasets: IRCAD (liver CT images), Bullitt (brain MR
images) and the synthetic tool VascuSynth.
Their performances were investigated in various areas: in the whole organ, around the vascular network and at the bifurcations, together with a multiscale approach guided by the diameters of the vessels. Fig.~\ref{fig:vesselCCOheart} compares the behavior of the seven selected filters in a bifurcation area.
Two versions (based on different parameter optimisation strategies) were proposed, first in \citep{DBLP:conf/icpr/LamyMKPV20}, then in \citep{DBLP:journals/tmi/LamyMKP22}.
An open source version\footnote{\url{https://github.com/JonasLamy/LiverVesselness}}  of the code and a demonstrator\footnote{\url{https://kerautret.github.io/LiverVesselnessIPOLDemo/}} were made available in \citep{DBLP:conf/rrpr/LamyKMP21}, in order to allow any interested user to assess the performance of his/her own filters on any angiographic data. Note that using the online demonstration platform, users can freely test the filters on their own volume they can upload to the server.
Fig.~\ref{fig:bench_module2} illustrates a typical example of experiment generated from the proposed filtering benchmark that can be customized with various metrics (mean/std MCC or Dice), and on different region domains (organ, vessel neighborhood).
Different representations can be generated (histogram (a), ROC curve (b), and raw data results).
An example of results from experiments carried out on 8 methods is given in Tab.~\ref{tab:VesselssizeIrcad} for different regions of the vascular network.

Section~\ref{SSEC:Softwares} also presents an open-source software proposing all those filters into a single 3D Slicer plug-in.

\subsection{Angiographic Image Segmentation}
\label{SSEC:Angiographic Image Segmentation}

\begin{figure}[t!]
\centering
\includegraphics[width=\linewidth, clip=true, trim=20 20 20 20]{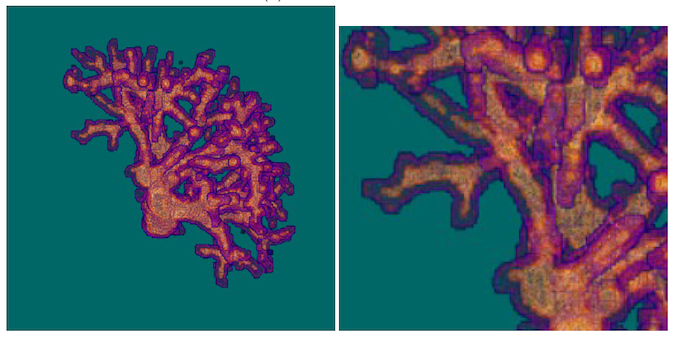}
\caption{Illustration of liver segmentation obtained with pre-filtered CT images and a 3D Dense U-Net model (left: global view; right: magnified view).}
\label{fig:vessel_seg}
\end{figure}

The segmentation of vessels in medical images has been covered by various methodologies over the last decades.
Many recent ones rely on deep learning models derived from the U-Net \citep{DBLP:conf/miccai/RonnebergerFB15}. 
The work conducted by Abir Affane \citep{DBLP:phd/hal/Affane22} during her PhD was first to adapt and compare 3D models to liver vessel segmentation, namely 3D U-Net, Dense U-Net, MultiRes U-Net on the IRCAD dataset \citep{Affane:JAS:2021}.
In addition,  different image structures were tested as inputs of the neural networks, with complete volumes, slabs (groups of slices) and patches.
It followed from this first study that the Dense U-Net approach, trained on image volumes, constitutes the best compromise in terms of performance vs. number of parameters. 
In a second time, a novel approach was designed, by leveraging the developments of the two PhD theses.
Indeed, as exposed in Table~\ref{tab:vessel_seg}, thanks to vesselness filters, it was possible to train deep learning models from vascular patterns provided by these operators, alternatively to native angiographic images \citep{Affane:IMU:2022}.
By comparing different filters, it was observed that the vesselness filter proposed by \cite{Jerman:TMI:2016}, used as a pre-processing step on CT images from IRCAD with a Dense U-Net model, was an accurate solution for hepatic vascular segmentation, as illustrated in Fig.~\ref{fig:vessel_seg}. By applying a few post-processing operators, we were able to obtain faithful segmentations, with a satisfactory preservation of bifurcations and small vessels. 
\begin{table}[htbp]
\centering
\small
\caption{Dice scores obtained with different U-Net-based models applied on the IRCAD dataset, with and without pre-processing filters. For each model, the best segmentation quality is depicted in bold face. More details can be found in~\citep{DBLP:phd/hal/Affane22,Affane:IMU:2022}.}
\begin{tabular}{l|lll}
\hline
\bf Preprocessing & \bf U-Net & \bf Dense U-Net & \bf MultiRes U-Net \\
\hline
Without & 0.586 & 0.671 & 0.694 \\
Jerman filter & \bf 0.712 & \bf 0.856 & \bf 0.835 \\
RORPO filter & 0.236 & 0.271 & 0.334 \\
Sato filter & 0.512 & 0.848 & 0.591 \\
Zhang filter & 0.201 & 0.311 & 0.212 \\
\hline
\end{tabular}
\label{tab:vessel_seg}
\end{table}

A major issue in medical image segmentation by deep learning models is the annotation of volumes by clinical experts. We present later in Section~\ref{SSEC:Softwares}  a 3D Slicer plug-in dedicated to this task for liver anatomy from CT or MR images. 

\subsection{Vascular Model Generation}
\label{SSEC:Vascular Model Generation}

Despite being composed of simple tubular elements, vascular networks are highly complex structures.
This complexity derives from their branching topology, their tortuous geometry and their highly multiscale organization, that goes from macroscopic vessels to microscopic ones.
It follows that angiographic imaging devices are not able to capture the complete information related to vascular networks.
Due to these limitations, some research efforts were dedicated to model the vessels, based on anatomical, physiological and biomechanical hypotheses. 
This led to numerical approaches for vascular model generation.
In particular, the CCO (Constrained Constructive Optimization) scheme was proposed by \cite{Schreiner1993}, first in 2 dimensions and then by \cite{Karch:CBM:1999} in 3 dimensions.
The CCO algorithm incrementally builds a binary tree by adding at each iteration a new terminal point.
Beyond a simple topological process, the strength of CCO lies in the spatial embedding of this tree, that defines a position for each branching point and a size for each segment based on physiologically-compliant rules guiding a spatially constrained optimization process.

\begin{figure}[t!]
  \begin{tabular}{cc}
    \includegraphics[ trim = 3.5cm 1.5cm 3.5cm 1.5cm, clip = true ,width=0.23\textwidth]{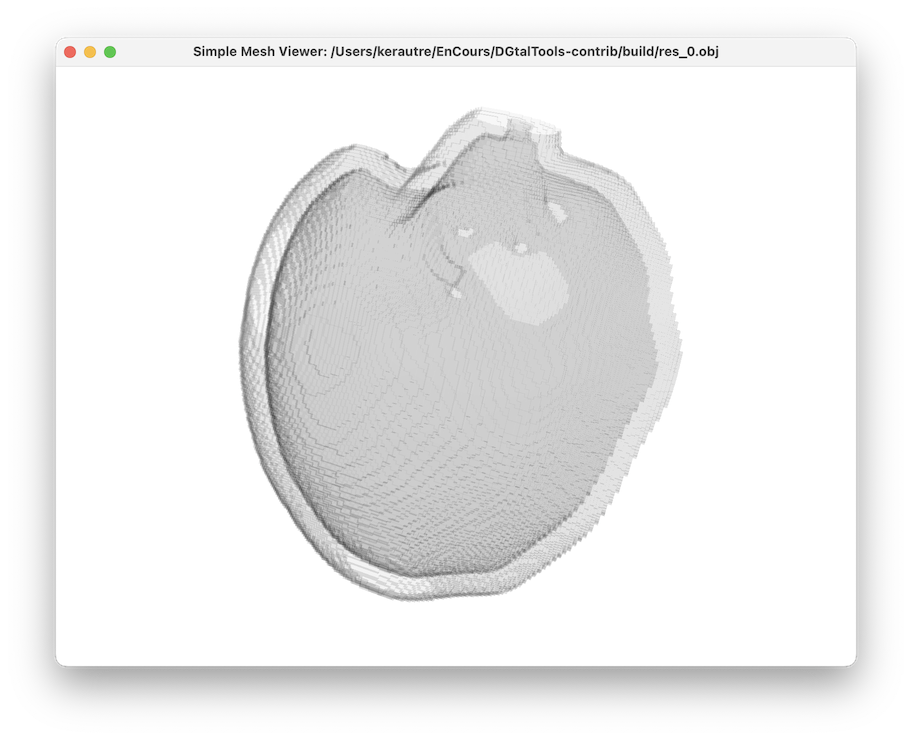}&
    \includegraphics[trim = 3.5cm 1.5cm 3.5cm 1.5cm, clip = true, width=0.23\textwidth]{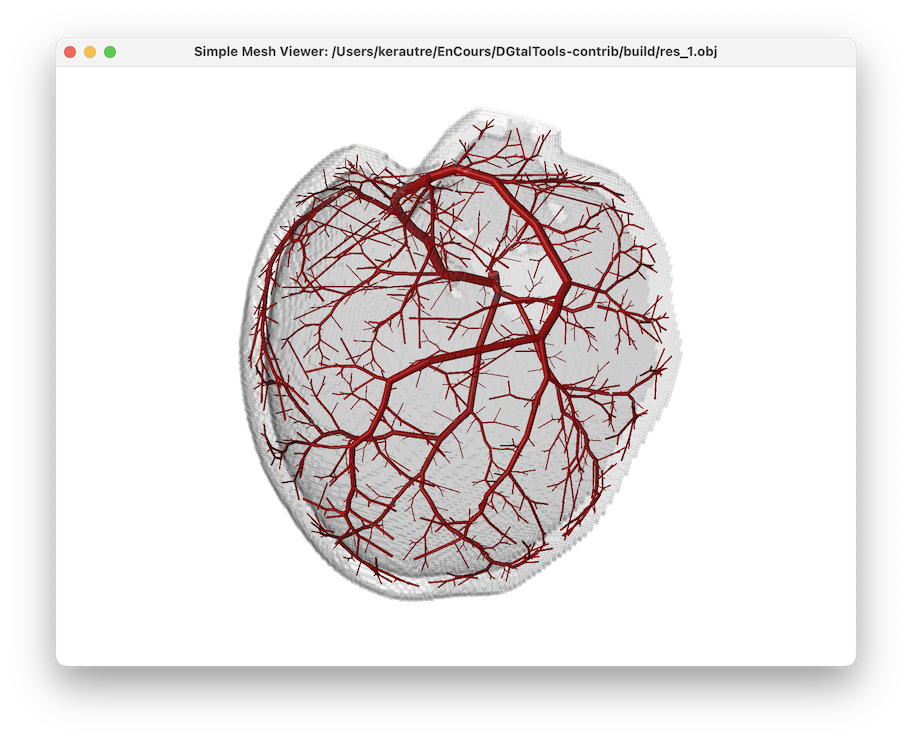}\\    
(a) & (b) 
  \end{tabular}
\caption{Generation of heart vascular network with OpenCCO, by
  taking into account the topology and geometry of the organ
  muscle. The generation domain was restricted to the heart wall
  illustrated on image (a) and the resulting vascular network is given
  on (b).  }
\label{fig:vessel_genHeart}
\end{figure}
\begin{figure*}[htbp]
\centering
  \begin{tabular}{cc}
  \includegraphics[width=.4\linewidth]{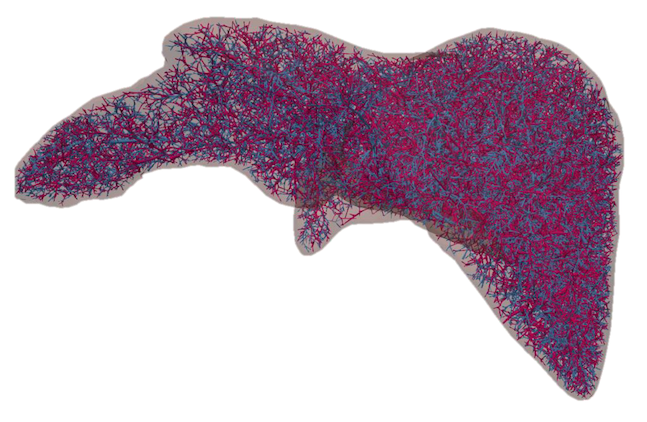}&
  \includegraphics[width=.45\linewidth]{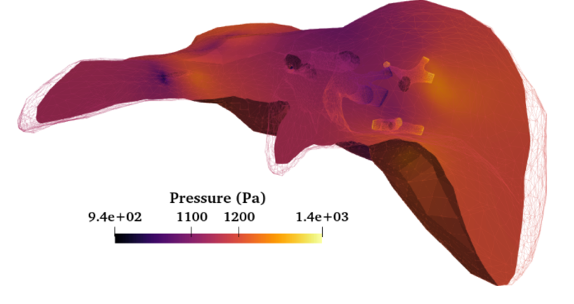}\\
  (a)  & (b) 
\end{tabular}
\caption{Generation of liver vessels by means of OpenCCO (a), considering branches of portal and hepatic veins as initialization of the algorithm. Computation of the pressure in the liver parenchyma (b) thanks to our numerical simulation pipeline.}
\label{fig:vessel_genLiver}
\end{figure*}
This seminal work led to the further development of variants, including the popular VascuSynth \citep{hamarneh:vascusynth:2010} tool and non-open implementations of CCO, \emph{e.g.} by \cite{Jaquet:TBE:2019}.
Motivated by research reproducibility with open software dissemination, and convinced by the relevance of CCO as a way to relevantly complete the information missed by angiographic images, we revisited the CCO algorithm and we proposed an open-source implementation together with a demonstrator\footnote{\url{https://opencco-team.github.io/OpenCCO-IPOL_Demo}} described by \cite{DBLP:journals/ipol/KerautretNPTJ23}.
Fig.~\ref{fig:vessel_genHeart} illustrates a result obtained from the online demonstration where the user defines a restricted domain defined on the heart wall.
Revisiting this algorithm led us to investigate some tricky parts of the optimization process, in particular the non-linear equation system governing the progressive construction of the vascular tree, initially solved as proposed by \cite{Kamiya1972OptimalBS}.
These efforts opened the way to new efficient optimization strategies, such as proposed by \cite{DBLP:conf/dgmm/ChouSNPTV24} and paving the way towards improved vascular tree modeling processes.

In section~\ref{SSEC:Vascular Simulation}, we expose how we employed this software in order to model the liver vasculature for the simulation of hepatic vascular functioning. 

\subsection{Vascular Simulation}
\label{SSEC:Vascular Simulation}

We finally  addressed the challenge of numerical simulation of liver perfusion, by taking into account vessels from medical images. This is a very hard task, due to the multi-scale high complexity of the liver functioning, from hepatic lobules to fine then 
 larger vessels, and to the partial representation of liver anatomy acquired from medical images~\citep{DBLP:conf/miccai/LebreAVLGBVMAC17}. During the post-doc fellowship carried out by Mohamed A. Chetoui, and to represent a large volume of the liver vasculature with the finest vessels, we used the aforementioned OpenCCO tool, as illustrated in Fig.~\ref{fig:vessel_genLiver}~(a). 
We generated this vascular network starting from the larger branches of hepatic and portal veins obtained from medical images. The liver volume boundary was also extracted from such images. Then, we developed a novel computing method to obtain physiological parameters (pressure, flow, permeability) of hepatic perfusion~\citep{Chetoui23}. We opted for a multi-compartment approach~\citep{DBLP:journals/corr/RohanLJ16}, and by coupling Darcy law and computational fluid dynamics, in order to calculate these parameters inside the liver parenchyma and vessels. An example of result obtained with our system is depicted in Fig.~\ref{fig:vessel_genLiver}~(b), where we show the pressure calculated within the liver volume. We tested our approach on 1D and 3D configurations to validate the correct behavior of the simulation regarding a \emph{in-silico} realistic liver physiology model. Moreover, we developed this complete pipeline into a single Python framework by using the FreeFEM library~\citep{MR3043640}.

\subsection{Software developments}
\label{SSEC:Softwares}

During the project, we developed and disseminated two official open-source 3D Slicer~\citep{Kikinis2014} plug-ins.  

\texttt{SlicerRVXVesselnessFilters}\footnote{\url{https://github.com/R-Vessel-X/SlicerRVXVesselnessFilters}} gathers the aforementioned vesselness filters (Section~\ref{SSEC:Angiographic Image Filtering}) into a single framework, which facilitates their use for biomedical and medical image processing in 3D Slicer. Any user can set the parameters related to each algorithm, and interactively visualize the result. 

\texttt{SlicerRVXLiverSegmentation}\footnote{\url{https://github.com/R-Vessel-X/SlicerRVXLiverSegmentation}}  aims at interactively aiding  clinical experts and biomedical engineers in liver anatomy segmentation from medical images (\emph{e.g.} CT, MRI), comprising the parenchyma, hepatic and portal vessels and potential tumors. We shew in~\citep{DBLP:journals/jossw/LamyPLMKPFV22} that this tool enables a faster annotation compared to commercial solutions, mostly for MRI data. It can thus be applied in medical analysis contexts, as we did \emph{e.g.} for registration ~\citep{DBLP:conf/ipta/DebrouxLMGV20}. Moreover, this is the very first time that such plug-in incorporates deep models based on MONAI~\citep{DBLP:journals/corr/abs-2211-02701} for automatic liver volume segmentation.

\section{Concluding Remarks and Perspectives}
\label{SEC:Concluding Remarks and Perspectives}

The R-Vessel-X project provided extensive research outcomes, covering various topics related to 3D angiographic image analysis, such as filtering, segmentation, modeling and simulation. We also developed softwares so that researchers, medical doctors and engineers can use essential vascular tools in their daily practice. The \texttt{SlicerRVXLiverSegmentation} plug-in is used in the radiology department at CHU (University Hospital) of Clermont-Ferrand, which enables clinical research in cardio-vascular pathologies. 
As perspective, the combination of these independent products would be beneficial for the community. Our goal is to propose new public and open-source plug-ins or softwares still related to angiographic imaging. Furthermore, we aim to continue our investigations on biomedical pre-clinical imaging~\citep{DBLP:conf/isbi/AlvarezPRV23}, and draw relations with human-scale vascular systems. 
This project also enabled new on-going researches, for example in deep vessel segmentation for the reconnection of vascular structures by an hybrid approach~\citep{DBLP:journals/ijon/EstevesVM24} or the combination of vesselness filters in deep U-Net models~\citep{Garret24}.

\section*{Compliance with Ethical Standards}
During the R-Vessel-X project, we used datasets and tools publicly available such as IRCAD~\citep{Soler2010}, Bullitt~\citep{Bullitt2005} or VascuSynth~\citep{hamarneh:vascusynth:2010} to obtain angiographic images. Also, each referenced article produced during the project met the requirements for ethical standards to get published. As a matter of reproducibility and availability of our developments, we created permanent links for our softwares SlicerRVXLiverSegmentation~\citep{DBLP:journals/jossw/LamyPLMKPFV22}, SlicerRVXVesselnessFilters~\citep{DBLP:journals/tmi/LamyMKP22} and OpenCCO~\citep{DBLP:journals/ipol/KerautretNPTJ23}.  

\section*{Acknowledgement}
The research leading to these results was funded by the French \emph{Agence Nationale de la Recherche} (Grant Agreement ANR-18-CE45-0018), which is gratefully acknowledged by the authors.

\bibliographystyle{model2-names.bst}\biboptions{authoryear}
\bibliography{biblio}

\end{document}